\documentclass[apj]{emulateapj}

\usepackage[T1]{fontenc}
\usepackage{amsmath,xspace}
\usepackage{amssymb}	
\usepackage[greek,english]{babel}
\usepackage{textcomp}
\usepackage{color}
\usepackage{footnote}

\usepackage{graphicx}	
\usepackage{gensymb}

\newcommand{\rmc}{RMC$\,$127\xspace}
\newcommand{\rmctre}{RMC$\,$143\xspace}
\newcommand{\suno}{S$\,$61\xspace}

\newcommand{\qso}{QSO$\,$J0635-7516\xspace}
\newcommand{\bpcal}{QSO$\,$J0538-4405\xspace}
\newcommand{\fluxcal}{Pictor A\xspace}

\shorttitle{RMC127 as seen with ALMA and ATCA}
\shortauthors{Agliozzo et al.}

\begin{document}


\title{The Luminous Blue Variable RMC127 
    as seen with ALMA and ATCA}


\email{c.agliozzo@gmail.com}



\author{C. Agliozzo\altaffilmark{1,2}, C. Trigilio\altaffilmark{3}, G. Pignata\altaffilmark{2,1}, N. M. Phillips\altaffilmark{4,5}, R. Nikutta\altaffilmark{6,7}, P. Leto\altaffilmark{3},  G. Umana\altaffilmark{3}, A. Ingallinera\altaffilmark{3},  C. Buemi\altaffilmark{3}, F. E. Bauer\altaffilmark{7,1,8},  R. Paladini\altaffilmark{9}, A. Noriega-Crespo\altaffilmark{10}, J. L. Prieto\altaffilmark{11,1}, M. Massardi\altaffilmark{12},  L. Cerrigone\altaffilmark{13}}


\altaffiltext{1}{Millennium Institute of Astrophysics (MAS), Nuncio Monse{\~{n}}or S{\'{o}}tero Sanz 100, Providencia, Santiago, Chile}
\altaffiltext{2}{Departamento de Ciencias Fisicas, Universidad Andres Bello,  
Avda. Republica 252, Santiago, 8320000, Chile}
\altaffiltext{3}{INAF-Osservatorio Astrofisico di Catania, Via S. Sofia 78, I-95123 Catania Italy}
\altaffiltext{4}{European Southern Observatory, Alonso de C\'{o}rdova 3107, Vitacura, Santiago, Chile}
\altaffiltext{5}{Joint ALMA Observatory, Alonso de C\'{o}rdova 3107, Vitacura, Santiago, Chile}
\altaffiltext{6}{National Optical Astronomy Observatory, 950 N Cherry Ave, Tucson, AZ 85719, USA}
\altaffiltext{7}{Instituto de Astrof{\'{\i}}sica and Centro de Astroingenier{\'{\i}}a, Facultad de F{\'{i}}sica, Pontificia Universidad Cat{\'{o}}lica de Chile, Casilla 306, Santiago 22, Chile}
\altaffiltext{8}{Space Science Institute, 4750 Walnut Street, Suite 205, Boulder, Colorado 80301} 
\altaffiltext{9}{Infrared Processing Analysis Center, California Institute of Technology, 770 South Wilson Ave., Pasadena, CA 91125, USA}
\altaffiltext{10}{Space Telescope Science Institute, Space Telescope Science Institute
3700 San Martin Dr., Baltimore, MD, 21218}
\altaffiltext{11}{N\'{u}cleo de Astronom{\'{\i}}a de la Facultad de Ingenier\'{i}a, Universidad Diego Portales, Av. Ej\~{e}rcito 441, Santiago, Chile}

\altaffiltext{12}{INAF-Istituto di Radioastronomia, via Gobetti 101, 40129, Bologna, Italy}
\altaffiltext{13}{ASTRON, Oude Hoogeveensedijk 4, 7991 PD, Dwingeloo,
The Netherlands}



\begin{abstract}
 We present ALMA and ATCA observations of
  the luminous blue variable \rmc. The radio maps show for the first
  time the core of the nebula and evidence that the nebula is strongly
  asymmetric with a Z-pattern shape. Hints of this morphology are also
  visible in the archival \emph{HST} $\rm H\alpha$ image, which
  overall resembles the radio emission. The emission mechanism in the
  outer nebula is optically thin free-free in the radio. At
  high frequencies, a component of point-source emission appears at
  the position of the star, up to the ALMA frequencies.  The rising
  flux density distribution ($S_{\nu}\sim \nu^{0.78\pm0.05}$) of this
  object suggests thermal emission from the ionized stellar wind and
  indicates a departure from spherical symmetry with
  $n_{e}(r)\propto r^{-2}$. We examine different scenarios to explain
  this excess of thermal emission from the wind and show that this can
  arise from a bipolar outflow, supporting the suggestion by other
  authors that the stellar wind of \rmc is aspherical. We fit the data
  with two collimated ionized wind models and we find that the
  mass-loss rate can be a factor of two or more smaller than in the
  spherical case. We also fit the photometry obtained by IR space
  telescopes and deduce that the mid- to far-IR emission must arise
  from extended, cool ($\sim80\,\rm K$) dust within the outer ionized
  nebula. 
  Finally we discuss two possible scenarios for the nebular morphology: the canonical single star expanding shell geometry, and a precessing jet model assuming presence of a companion star.

\end{abstract}






\section{Introduction}
It is widely accepted that the final destiny of a massive star is
ruled by the mass-loss suffered during its post-Main Sequence (MS)
evolution, and by how much mass remains at its death. For instance,
the earliest O-type stars have to rapidly lose their hydrogen envelope
(a few to a few tens of solar masses) in order to turn into Wolf-Rayet
(WR) stars. The transition between the MS and the WR phase must be
short, of the order of $10^{4}-10^{5}\,\rm yr$. Enhanced mass-loss is
needed to reduce the envelope mass, through line-driven stellar winds
or short-duration eruptions \citep[e.g.][]{1994HD, 2006S&O}. The stars
with the highest known mass-loss rates
($\dot{M}\gtrsim10^{-5}\,\rm M_{\odot}\,yr^{-1}$) are the Luminous
Blue Variable (LBV) stars, so called due to their location in the H-R
diagram and because they show spectroscopic and photometric
variability during a period of enhanced mass-loss caused by
instabilities, as reviewed in \citet{1994HD}. These instabilities have
yet to be conclusively explained, but several physical mechanisms have
been proposed: vicinity to the (modified) Eddington limit due to an
excess of radiation pressure; hydrodynamic (sub-photospheric)
instabilities; rapid rotation and/or close-binary systems.

\citet{2015S&T} noticed that the known Galactic and Magellanic LBVs tend to be
isolated from massive star clusters. 
Hence they challenged the traditional single-star view of LBVs,
proposing that the LBV phenomenon (strong instabilities and enhanced
mass-loss) is instead due to interacting binaries, with a ``mass
donor'' (e.g. WR star) and a ``mass gainer'' (LBV). The
mass-transfer would ``rejuvenate'' the LBV star, whose evolution, as a
consequence, would bifurcate from that of the other stars in the
cluster where it formed. More recently, \citet{2016Humphreys} tested the same analysis for the LBVs in M31 and M33 and they also removed ``seven stars with no clear relation to LBVs'' from the sample of \citet{2015S&T}. \citet{2016Humphreys}
then found that the LBVs distribute similarly to their O-type sisters or
to the Red Supergiant (RSG) ones, depending on their initial
mass and evolutionary state. Therefore, they revived the scenario for the evolution of a 
single massive star that approaches the Eddington limit. 

The picture is still unclear and, due to the rarity of these objects,
together with the rapid evolution of massive stars, we are still
unable to put together all the pieces of the puzzle. On one hand it
has been accepted that some LBVs and Ofpe \citep{1989BW} super-giants
are physically related, with the latter considered the quiescent state
of a massive (O-type) LBV \citep[e.g.][]{1986StahlO}. On the other
hand, there is no evidence of their relationship with other massive
stars, despite some suggestions: for instance, supergiant B[e] stars
\citep[][and following studies]{1985Zickgraf} and Of?p stars
\citep[the question mark indicates doubt that these stars are normal
Of super-giants,][]{1972Walborn, 1977Walborn}. The B[e] supergiants
are fast rotators and possess a dense and slow disk in their
equatorial plane and a faster outflow along the polar axis. The Of?p
stars have been found to be oblique magnetic rotators \citep[][and
ref.  therein]{2015Walborn}. Interestingly, the Galactic LBV AG
Carinae (AG Car) has been found to be a fast rotator and its projected
rotational velocity has been seen to change during LBV variability
\citep{2006Groh}, but magnetic fields have not been detected in any
known LBV.

The distinct morphologies observed in the nebulae around some
candidate and confirmed LBVs, formed as a consequence of the intense
mass-loss, suggest different shaping mechanisms \citep{1995Nota}. The
morphologies of some nebulae are consistent with a symmetric mass-loss
\citep[e.g. Gal 79.29+0.46, \suno,][]{1994Higgs, 2003Weis, 2011UmanaB,
  2012Agliozzo, 2014Agliozzo, 2017AgliozzoA}. However, the majority of
the observed nebulae have a bipolar morphology
\citep[e.g.][]{2011Weis} indicating aspherical mass-loss
\citep[e.g. Gal 026.47+0.02, ][]{2012Umana} or an external shaping
mechanism \citep[e.g. IRAS 18576+0341, HR Car,][]{2005Umana,
  2010Buemi, 2017Buemi}. Usually, bipolar or equatorial mass-losses
have been proposed. Departure from spherical-symmetry has been
directly observed in the winds of AG Car, HR Car and \rmc
\citep[e.g.][]{1994Leitherer, 1995Clampin, 1993Schulte}, but whether
aspherical mass-loss is an intrinsic property of LBVs has not been
established \citep[e.g.][]{2005Davies}. To explain LBVs with bi-polar
or ring nebulae, enhancement of mass-loss in the equatorial plane of
the star has often been invoked, the cause possibly being the fast
rotation of the star, or the presence of a companion star, or a
magnetic field \citep[e.g.][and ref. therein]{2015Gv}.

\rmc (HD 269858) is a well-known LBV located in the Large Magellanic
Cloud. In the last decades it has been observed in both quiescent and
active states, during which the stellar spectrum changed from Ofpe to
A spectral types, through intermediate types {B1--2}, B7 and B9. At
the beginning of the 2000s it began its decline towards the quiescent
state \citep{1977Walborn,1982Walborn,1983Stahl,1988Wolf,
  2008Walborn}. The first high-resolution image \citep{1993Clampin}
revealed the presence of a ``diamond-shaped nebula'' associated with
the star.  By means of spectro-polarimetric studies in the optical,
\citet{1993Schulte} found a high degree of polarization at the
position of the star, similar to B[e] stars. They proposed two
geometries for the stellar wind and the aligned outer optical nebula
around \rmc: a highly inclined bi-polar nebula or a disk or ring of
material seen edge-on. This polarization was later confirmed by
\citet{2005Davies}. \citet{1998Smith} studied the kinematics of the
nebula and they interpreted the data as two expanding shells (with the
inner one $\lesssim$ 0.6 pc from the star). Later \citet{2003Weis}
obtained high resolution images in $\rm H\alpha$ with the Wide Field
Planetary Camera 2 (WFPC2) on board the \emph{Hubble Space Telescope}
(\emph{HST}). The authors described the nebula as
comprising two Eastern and Western
\emph{Rims} and two Northern and Southern \emph{Caps}. With a kinematic
analysis of spectroscopic observations, they also reported that the
Northern and Southern regions are blue- and red-shifted, respectively,
with respect to the center of motion. Finally, RMC127 is a fast
rotator, with a projected rotational velocity of
$\sim105\,\rm km\,s^{-1}$ \citep{2017AgliozzoB}.

In this paper we present a new dataset, acquired with ALMA and ATCA. 
We discuss the detection of a compact object associated with the
central star from the centimeter (with ATCA) to the sub-millimeter
(with ALMA). To
understand the nature of this object, we complement the radio and
sub-mm data with the IR photometry from space telescopes extracted from the public catalogs. Together with the central
object, we also analyze the outer nebula that we detected with ATCA at
17 GHz. In this analysis we also include the maps at 5.5 and 9 GHz
that were presented in \citet[][hereafter Paper I]{2012Agliozzo}. 

The paper is organized as follows: we present our dataset in Section~\ref{sec:obs}. In Section~\ref{sec:anal} we analyze 
the radio and sub-mm emission, including the IR photometry from space telescopes. In Section~\ref{sec:discussion} we discuss the central object emission and fit it with two \citet{1986Reynolds} models
for collimated ionized winds. In Section \ref{sec:outernebula} we comment on different scenarios for the geometry of the outer nebula. Finally we summarize our results in Section \ref{sec:summary}.

\section{Observations and data reduction}
\label{sec:obs}
\begin{figure*}
  \centering
  \includegraphics[width=.8\linewidth]{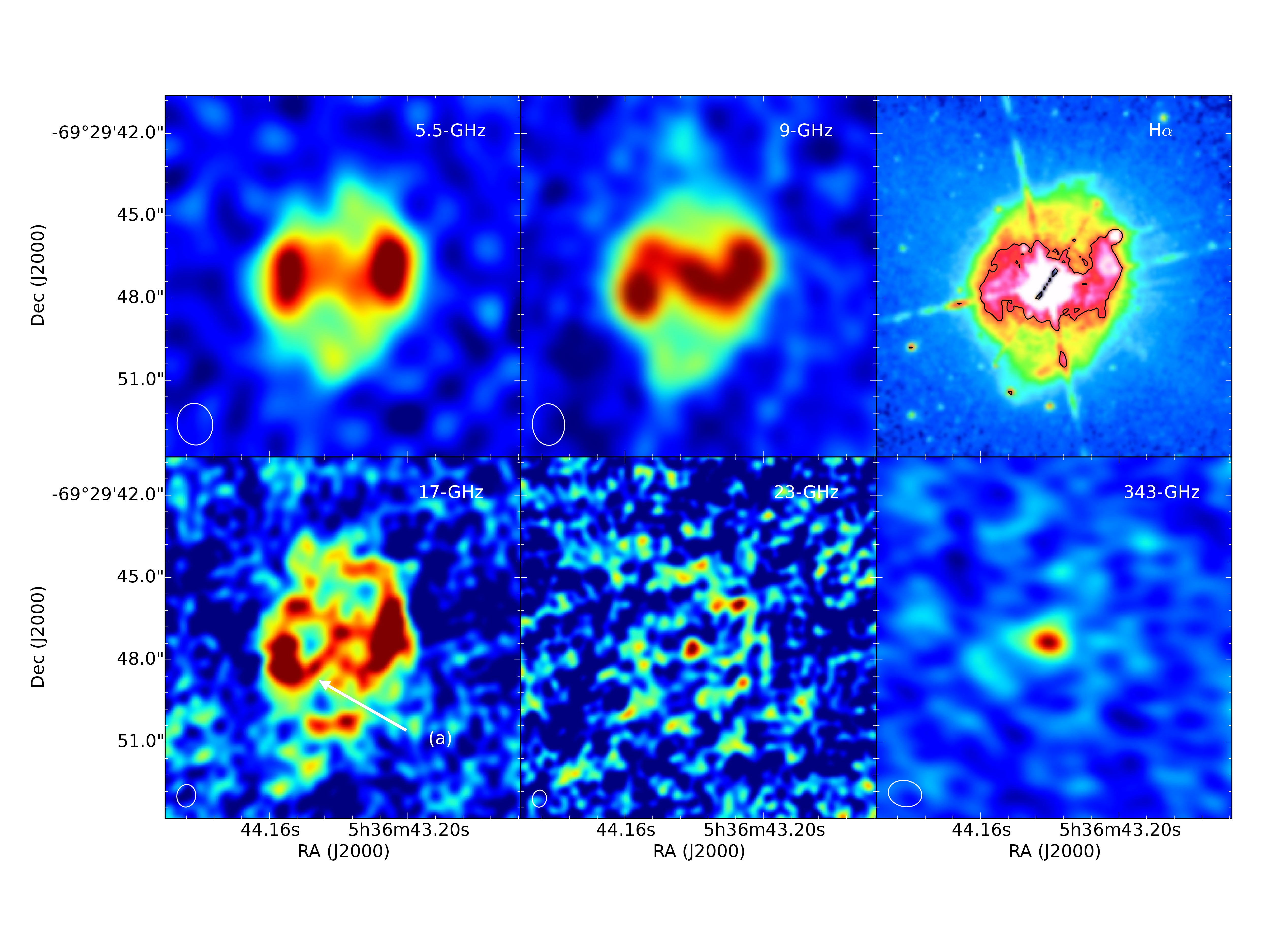}
  \caption{From top-left to top-right: ATCA 5.5 and 9$\,$GHz, and
    \emph{HST} archival F656N $\rm H\alpha$ image. From bottom-left to
    bottom-right: ATCA 17 and 23$\,$GHz, and ALMA 343.5$\,$GHz.  All
    images have the same field size. In each map, the synthesized beam
    is indicated with a white ellipse. At the lowest frequencies the
    emission from ionized gas in the nebula is detected and is
    co-spatial with the optical emission seen in $\rm H\alpha$. At the
    higher radio and sub-mm frequencies the central object is mostly
    the dominant component. In the optical image the black contour
    evidences hints of the bright Z-pattern shape visible, in
    particular, at 9 GHz. The white arrow and label (a) indicate the size of one diagonal arm (about 2.8 arcsec).}
  \label{fig:maps}                                                                                                                                                                                                                                                                                                            
\end{figure*}

\begin{table*}
  \centering
  \caption{Summary of observations and images properties.}
  \label{tab:obs}
  \begin{tabular}{lccccccc} 
    \tableline\tableline
    Date         & Array & $\nu$  & LAS & Synthesized Beam& PA & Peak & noise \\
                 &       & (GHz)  & (arcsec) &  FWHM (arcsec)& (deg) & (mJy beam$^{-1}$) & (mJy beam$^{-1}$) \\
    \tableline
    2011-04-18/20& ATCA  & 5.5 & 22.0 & 1.53$\times$1.29&8.3&0.41&0.02\\
    2011-04-18/20& ATCA  & 9 & 12.3 &  1.52$\times$1.17&3.4&0.40&0.03\\
    2012-01-21/23& ATCA  & 17 & 6.5 &  0.82$\times$0.69&-8.0&0.17&0.02\\
    2012-01-21/23& ATCA  & 23 & 4.1 &  0.62$\times$0.51&-8.0&0.16&0.03\\
    2014-12-26   & ALMA-LSB\footnote{Lower sideband map.}&337.5 & 9 &  1.26$\times$0.97&78.3&1.130&0.095\\
    2014-12-26   & ALMA\footnote{All the bandwidth. }&343.5 & 9 &  1.23$\times$0.95&78.4&1.140&0.072\\
    2014-12-26   & ALMA-USB\footnote{Upper sideband map. }&349.5 & 9 &  1.21$\times$0.93&78.6&1.18&0.11\\
    \tableline
  \end{tabular}
\end{table*}

\subsection{ALMA observation and data reduction}


\rmc ($\rm 05^{h} 36^{m} 43.688^{s}$
$\rm -69^{\circ} 29^{\prime} 47.52^{\prime\prime}$ ICRS) was observed
as part of an ALMA Cycle-2 project studying three Magellanic LBVs
(Project ID: 2013.1.00450.S), including LBV \rmctre (Agliozzo et al. in
prep) and candidate LBV \suno \citep[][hereafter Paper
II]{2017AgliozzoA}. The observations consisted of a single execution of
80 minutes total duration on 2014-12-26 with 40 12m antennas, with
projected baselines from 10 to 245 m. The integration time per target
was 16 minutes. A standard Band 7 continuum spectral setup was used,
providing four 2 GHz-width spectral windows of 128 channels of XX and
YY polarization correlations centered at approximately 336.5(LSB),
338.5(LSB), 348.5(USB) and 350.5(USB) GHz. Antenna focus was
calibrated online in an immediately preceding execution, and antenna
pointing was calibrated on each calibrator source during the execution
(all using Band 7). Scans at the science target tuning on bright
quasar calibrators \bpcal and \fluxcal (PKS J0519-4546; an ALMA
secondary flux calibrator `grid' source) were used for interferometric
bandpass and absolute flux scale calibration. Astronomical calibration
of complex gain variation was made using quasar
 \qso interleaved with observations on the
science targets approximately every six minutes.

As already mentioned in Paper II, atmospheric conditions were marginal for the combination of frequency and necessarily high airmass (transit elevation $43^\circ$ for \rmc), requiring extra calibration steps in order to minimize image degradation due to phase smearing, to provide correct flux calibration and to maximize sensitivity by allowing inclusion of shadowed antennas. Details are presented in the Appendix, as they are expected to be of use for improving calibration for similar ALMA observations in marginal weather conditions at high airmass and/or with significant airmass difference between targets and gain calibrator (\qso), especially at bands 7 and above. This was performed in collaboration with staff at the Joint ALMA Observatory (JAO) who are working on these aspects of calibration.

In this work we used intensity images produced from naturally weighted visibilities in order to maximize sensitivity and image quality (minimize the impact of phase errors on the longer baselines). We imaged all spectral windows together ($343.5\,{\rm GHz}$ average; approximately $7.5\,{\rm GHz}$ usable bandwidth), obtaining an RMS noise of $0.072\,{\rm mJy}\,{\rm beam^{-1}}$ in the image. The proposed sensitivity of $40\,\mu{\rm Jy}\,{\rm beam^{-1}}$ could not be achieved as no further executions were possible during
the appropriate array configuration in Cycles 2 and 3. Therefore, the nebula was not detected. The central object was instead detected at $15 \sigma$.

We separately imaged the pairs of spectral windows in the two receiver sidebands ($337.5$ and $349.5\,{\rm GHz}$ centers; approximately $3.75\,{\rm GHz}$ bandwidth each), yielding RMS noise of $\sim\!0.10\,{\rm mJy}\,{\rm beam^{-1}}$ in the images. The two sideband images were used for an internal measure of the spectral index, and as a cross-check on the data quality. Fig.~\ref{fig:maps} illustrates the full bandwidth (7.5 GHz) map at 343.5 GHz. The synthesized beam is approximately $1.23''\times0.95''$. Table~\ref{tab:obs} lists details of the observations and of the resulting images, including date of observations, interferometer, central frequency, largest angular scale (LAS), synthesized beam size (FWHM) and its position angle (PA), peak flux density on nebula and noise in the residual maps. Flux calibration uncertainty is estimated as 5\% ($1\sigma$), although the peak flux may have a small  additional systematic error to lower value due to residual phase smearing at a similar level. These uncertainties will be strongly correlated between the three images (average, LSB and USB) due to the small fractional frequency differences ($3.5$\% between sideband centers) and minimal differences in atmospheric transmission, so the measurement of spectral index from only the ALMA data should be entirely noise-limited. Calibration and imaging of the ALMA data were performed using the Common Astronomy Software Applications (CASA) package, version 4.5 \citep{2007McMullin}. 

\subsection{ATCA data}

The 17 and 23-GHz ATCA data of \rmc were obtained as part of
observations of three magellanic LBV nebulae (Project ID: C1973),
including \rmctre and \suno. The observations were performed by using the
array in the extended configuration (6-km) and the Compact Array
Broadband Backend (CABB) receiver ``15-mm Band'' (16-25 GHz) during
2012 (see Table~\ref{tab:obs}).  We used observations of QSO
J1924$-$2914, ICRF J193925.0$-$634245 and ICRF J052930.0$-$724528 for
determining the bandpass, flux density and complex gain solutions,
respectively. We also determined the polarization leakages and applied
the solutions to the data in order to calibrate the cross-hands
visibilities. Once corrected, the visibilities were inverted by
Fourier transform.  We imaged the I, V, Q and U Stokes parameters with
a natural weighting for \emph{uv}-data, for the best sensitivity. In
the intensity map, we detected the nebula and the central object (see
Fig.~\ref{fig:maps}). We did not detect any polarization, as expected
due to the low dynamic range.  At 23 GHz there are positional errors
in declination which are $\sim 0.4$ arcsec, almost half a synthesized
beam. A source of error can be the fact that the phase calibrator was
systematically at lower elevations than the science target. Given the
poor weather and the proximity to the 22 GHz water line, systematic
errors in the estimation of the atmospheric path length may contribute
to this astrometric error.  More information on these observations and
data reduction can be found in the paper dedicated to \suno (Paper II).

We also re-analyze the 5.5 and 9-GHz data from the ATCA observations
performed in 2011 by using the CABB ``4cm-Band'' (4-10.8 GHz)
receiver. These data were presented in Paper I. Here we reprocessed the
data at 5.5 GHz by using a Briggs weighting scheme, with parameter
robust=0, in order to match the angular resolution at 9-GHz. The ATCA
maps are shown in Fig.~\ref{fig:maps} together with the ALMA
one. 

\subsection{VISIR observations}
We also observed \rmc at the Very Large Telescope (VLT) with the
instrument VISIR, as part of a project including \rmctre and \suno
(Project ID: 095.D-0433). An observing block in the narrow bandwidth
filter PAH2$\_$2 (centered at 11.88 $\rm \mu$m) was successfully
executed on 2015-11-01. The source was observed at an airmass of 1.6
and the seeing in the V-band was not better than 1 arcsec. The
observing mode was set for regular imaging, with pixel scale of 0.045
arcsec. Four perpendicular nodding positions were used. \rmc was not
detected, in part because the data were acquired for only $20\%$ of
the desired total exposure time. According to the VISIR Exposure
Time Calculator, for a point source of $40\,\rm mJy$ the
signal-to-noise (S/N) would be $<3\sigma$ in each
sub-images. Therefore the source could not be detected. We do not show
the VISIR image because it is just noise, without detectable sources
(the field of view is only $38\times38$ arcsec$^{2}$).

\subsection{{\it Hubble Space Telescope\/} archival data}
We retrieved from
the STScI data archive the $\rm H \rm \alpha$ {\it Hubble Space Telescope\/} ({\it HST\/}) images of \rmc (Project ID: 6540), published by \citet{2003Weis}. The images were obtained with
the Wide Field and Planetary Camera 2 (WFPC2) instrument using the $\rm H \rm \alpha$-equivalent filter F656N and
reduced by the standard {\it HST\/} pipeline. We reprocessed the data as described in Paper I and II. We obtained a final image that matches with \citet{2003Weis}.  
We show it in the top line of Fig. \ref{fig:maps}.

\section{Analysis and results}
\label{sec:anal}

\subsection{The ATCA and ALMA maps}
 \label{sec:morphology}
  \begin{figure}
 \centering
	\includegraphics[width=1\linewidth]{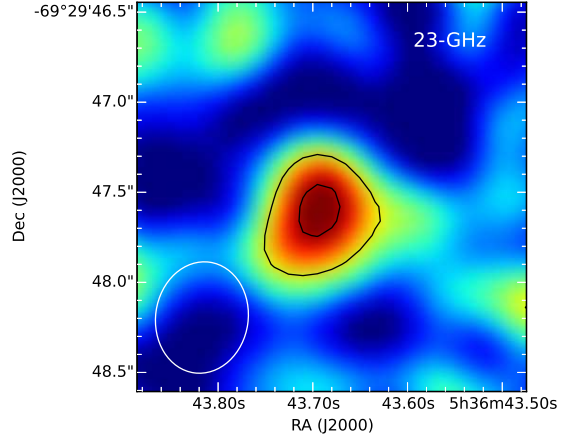}
    \caption{Zoom-in of the ATCA 23 GHz map, centered on the central
      object. The white ellipse in the corner represents the
      synthesized beam. The 23 GHz contour levels corresponding to the
      3 and 5$\sigma$ emission are plotted in black. The object
      appears slightly elongated, with a PA$\sim110^\circ$, similar to
      the PA of the polarized emission detected by
      \citet{1993Schulte}. However, we cannot exclude noise
      contamination at the 3$\sigma$ level.}
    \label{fig:zoom}                                                                                                                                                                                                                                                                                                            
\end{figure} 
 
\begin{table}
  \centering
  \caption{Central object peak flux density F$_{\nu}$ and nebula flux density S$_{\nu}$ at different frequencies.}
  \label{tab:fluxdensity}
  \begin{tabular}{ccccc} 
    \tableline\tableline
    $\nu$ & Synthesized Beam & PA    & F$_{\nu}$         & S$_{\nu}$ \\
    (GHz) & (arcsec)         & (deg) & (mJy beam$^{-1}$) & (mJy)\\
    \tableline
      5.5 & 1.53$\times$1.29 &  8.3  & $<0.255$          & 3.1$\pm$0.2\footnote{From Paper I}\\
        9 & 1.52$\times$1.17 &  3.4  & 0.08$\pm$0.05\footnote{Nebula flux subtracted.} & 3.3$\pm$0.4\footnote{From Paper I}\\
       17 & 0.82$\times$0.69 & -8.0  & 0.10$\pm$0.02     & 3.0$\pm$0.2\\
       23 & 0.62$\times$0.51 & -8.0  & 0.16$\pm$0.03     & --\\
    337.5 & 1.26$\times$0.97 & 78.3  & 1.130$\pm$0.095   & --\\
    349.5 & 1.21$\times$0.93 & 78.6  & 1.18$\pm$0.11     & --\\
    \tableline
  \end{tabular}
\end{table}

 In Fig.~\ref{fig:maps} we show the interferometric radio and sub-mm maps                                                                             
of \rmc and we compare them with the archival \emph{HST} $\rm H \alpha$ image, on the top-right. The
resolution of the radio maps corresponds to the synthesized beam and is
shown with white ellipses.

The radio maps reveal for the first time the inner part of the nebula. From low to high frequencies, different components dominate in the
distribution of the emission. At low frequencies the nebula (very likely ionized
by the central hot star) is the main source of radio emission, while
at higher frequencies the central object dominates the emission. When
the nebula is detected, the bipolar morphology is always evident. Previously, \citet{2003Weis} recognized in the \emph{HST} F656N $\rm H \alpha$ image an elongation culminating with the Northern and Southern caps. In the radio maps, in particular at 9 GHz, we notice an additional component at position angle (PA) $\sim70^\circ$, a bar or a ``diagonal arm''. This forms with the two Eastern and Western rims (``vertical arms'') a Z-pattern shape in the E-W direction. This is also visible in the \emph{HST} F656N $\rm H \alpha$ image (see black contour in the top-right panel), despite the spikes and artifacts (due to the bright central star) that affect the appearance of the nebular morphology. The radio maps present indeed a new insight in the core of the nebula. 

At 5.5 and 9 GHz the size of the nebula is approximately  $7\times6\,\rm arcsec^2$, or about $1.6\times1.4\,\rm pc^2$ assuming a distance of $48.5\,\rm kpc$ for the LMC. The measured size is consistent with the estimate determined from the optical image \citep{2003Weis}.  Since at 5.5 and 9 GHz the largest angular scale (LAS) is at least twice the size of the source (Table~\ref{tab:obs}), we do not expect any significant loss of flux at these frequencies due to the sampling of the $uv$ plane. 

At 17 GHz the source has the same extension and, roughly, the same morphology as at lower frequencies. However, the LAS is comparable to the source size; for this reason, even if the integrated flux density is preserved, artifacts can appear in the image.
We spatially integrated the flux density at 17 GHz. The new measurement together with the 5.5 and 9 GHz values (reported in Paper I) are listed in Table \ref{tab:fluxdensity}. They are all consistent with thermal free-free emission (see Sec. \ref{sec:spix}). In Table \ref{tab:fluxdensity} we also list the peak flux density at the central object position.

At 23 GHz the LAS is smaller than the source and, in fact, it is
possible to detect only the compact central object and the edges of
the nebula, while the extended flux is lost. At the ALMA frequency the
LAS is comparable to or larger than the size of the nebula as seen at
lower frequency. However, we only detect the compact central
source. The nebula around it is barely discernible. This is probably
caused by a low brightness of the source at the ALMA frequency
compared with the RMS of the map. Assuming that the nebula emits
through thermal optically thin free-free (Paper I), it is possible to
estimate the total flux of the nebula at 343.5 GHz by extrapolating the
measured flux density at low frequency with the typical power law
($S_{\nu}\sim \nu^{-0.1}$). The resulting flux density at 343.5 GHz is
$\sim2\,\rm mJy$. If we consider the number of ALMA beams in the
nebula, this flux density corresponds to an average brightness of
$0.14\,\rm mJy\,beam^{-1}$, which is only 2$\sigma$, therefore it was
not detected. Note that with the completion of observations of the
ALMA project (ID: 2013.1.00450S) it would have been possible to also detect
the free-free emission in the nebula. 

At
9-GHz the center of the nebula becomes as bright as the two rims. This is likely due to another emission component, 
which is partially resolved from the nebula at 17 GHz and also detected at 5$\sigma$ in the map at 23 GHz, where it appears as a compact
source (Fig.~\ref{fig:zoom}). This object has an apparent elongation in nearly E-W direction at the 3$\sigma$ level. The position angle (PA) of this object is $\sim110^\circ$, similar to the PA of the polarized emission detected by \citet{1993Schulte}. 
At the ALMA frequency the central object is the dominant emission
component. 

For the \hbox{23\,GHz} map, which is the highest-resolution map that
we obtained, we used the task \texttt{imfit} of CASA to fit a 2-D
Gaussian to the central object in the image-plane. The box region for
the fit was selected around the $\sim3\sigma$ contour level. The fit
results in a Gaussian with FWHM=$0.71\pm0.11$\,arcsec along the major
axis. The source is marginally resolved. The deconvolution from the
synthesized beam gives a size of 0.43 arcsec in RA, equivalent to
$\sim 3 \times10^{17}\,\rm cm$, or about 0.1 pc, at the distance of
48.5 kpc. However, at the 3$\sigma$ level the contours of this object
could be still confused with the noise in the map. Hence, the size
provided above must be considered an upper limit.

\subsection{Spectral index of the central object}
\label{sec:fluxdensity}

In Table~\ref{tab:obs} we report the peak flux density over the nebula. While at lower frequencies ($<$ 23 GHz) the peak
brightness is in the right arm (Western rim) of the nebula, at 23 GHz and between
337.5 and 349.5 GHz the peak of the emission is at the position of the
star. At these ALMA frequencies the peak flux density is almost three times higher than at
9 GHz (in the nebula), despite the ALMA synthesized beam being smaller.

The central object is not visible at 5.5 GHz. The reason can be
confusion with the nebula emission, due to poor resolution at this
frequency, combined with weaker emission. The latter possibility
suggests a rising flux density distribution (e.g.
$S_{\nu}\sim \nu^{\alpha}$), which is typical of stellar winds and
self-absorbed emission.  Note that the new map at 5.5 GHz has a
synthesized beam almost identical to that at 9 GHz. We extract the
peak flux density at 9 GHz ($0.36\pm0.03\,\rm mJy$) at the position of
the central object. Due to the low resolution this is contaminated by
the emission from the ``diagonal arm''. We then cut a slice along this
arm, and fitted a Gaussian to the brightness profile in the slice
along $\rm PA=70^\circ$. The distribution peak corresponds to a
brightness of 0.28 mJy and $\sigma$= 0.05 mJy. We subtract this value
from the peak flux density at the position of the central object and
derive a brightness of $0.08\pm0.05\,\rm mJy$ at 9 GHz. At higher
frequencies we will refer to the peak flux densities in
Table~\ref{tab:fluxdensity} extracted at the position of the central
object. Their associated errors are the noise $\sigma$ as estimated in
the residual maps (flux calibration errors are negligible).
\begin{figure}
  \centering
  \includegraphics[width=1\linewidth]{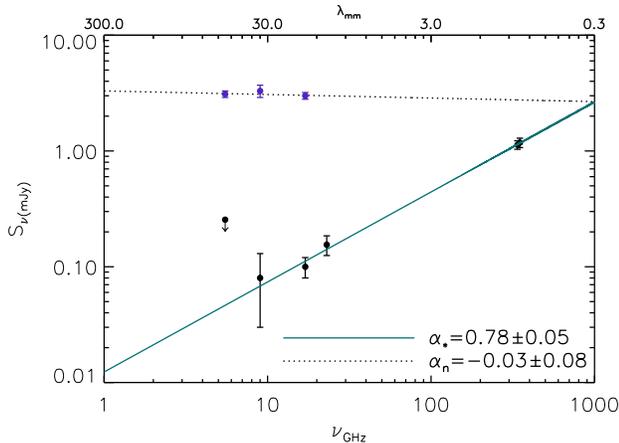}
  \caption{Black points: peak flux densities at 5.5 (upper limit), 9,
    17 and 23 GHz from the ATCA observations and at 337.5, 349.5 GHz
    from the ALMA observations, extracted at the position of the
    star. Solid line: weighted fit from 9 to 349.5 GHz. The positive
    slope suggests thermal emission from the ionized stellar wind. The
    spectral index $\alpha_{*}$ indicates departure from spherical
    symmetry with $n_{e}(r)\propto r^{-2}$. Purple points: spatially
    integrated flux densities of the nebula at 5.5, 9 and 17
    GHz. Dashed line: fit from 5.5 to 17 GHz, characterized by an
    $\alpha_{n}$ typical of optically thin free-free emission in the
    ionized
    nebula.}
  \label{fig:sed2}
\end{figure}

We derive a weighted fit of the power-law ($S_{\nu}\sim \nu^{\alpha}$) between the centimeter
and sub-millimeter flux densities of the central object (Fig.~\ref{fig:sed2}). This gives us a spectral index $\alpha$ 
of $0.78\pm0.05$, which is higher than the canonical value for ionized winds with spherical symmetry and $n_{e}(r)\propto r^{-2}$ \citep{1975Panagia,1975W&B}. Several mechanisms to explain the central object emission will be discussed in Section \ref{sec:discussion}. A potential caveat with the flux density distribution may be the presence of systematic errors in each individual measurement (Table~\ref{tab:fluxdensity} and Fig.~\ref{fig:sed2}) due to the differing beam sizes and the  nebular contributions to the extracted central object brightness in the maps. However, given the large frequency coverage (9 to 349.5 GHz) we are confident of the derived spectral index.

\begin{figure*}
\centering
\includegraphics[scale=0.35]{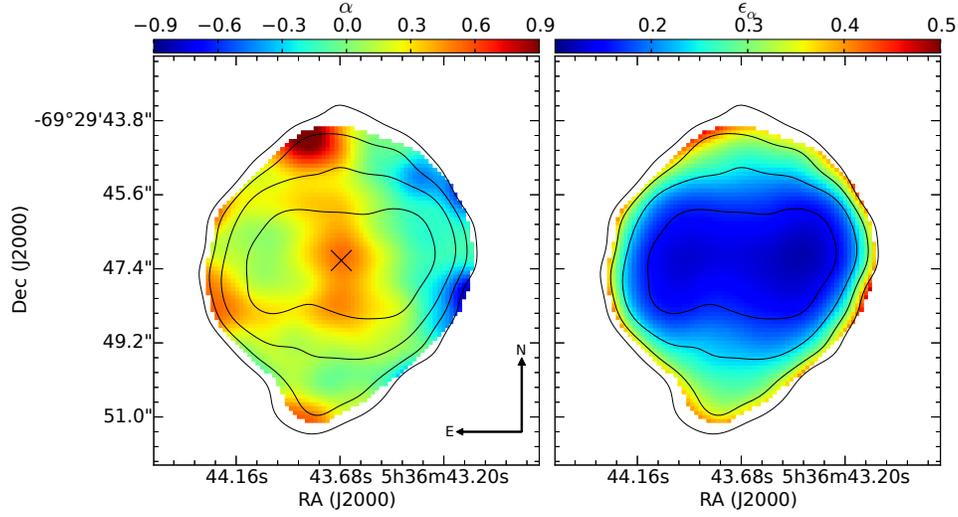}
\caption{
Left: Spectral index map between 9 and \hbox{5.5\,GHz}. Contours show the isophotes of the \hbox{9\,GHz} image at the resolution of the \hbox{5.5\,GHz} data. The central component has a spectral index consistent with a thermal wind \hbox{$\alpha\approx 0.6-0.7$}. The emission of the two arms in the West and East directions have spectral index consistent with optically thin free-free emission \hbox{$\alpha\approx 0$}.
Right: spectral index error map. The error associated with the spectral index map does non exceed 0.3 in most of the nebula. There is no evidence of non-thermal components.
}

\label{fig:spix_lbv}
\end{figure*}
\subsection{Spectral index of the outer nebula}
\label{sec:spix}

In Paper I we derived an average spectral index $\alpha \sim 0.1$ from
the spatially-integrated flux densities at 5.5 and 9 GHz. With the new
measurement at 17 GHz and the values from Paper I (see Table
\ref{tab:fluxdensity}), the average spectral index is $-0.03\pm0.08$
(Fig. \ref{fig:sed2}), typical of optically thin free-free emission,
with the flux density slightly decreasing at high frequencies and a
theoretical power law $\propto \nu^{-0.1}$. 

The images at radio wavelengths have angular resolution and
sensitivity that allow us to study any deviation from the typical
thermal free-free emission inside the nebula, by means of spectral index maps. 
The existence of
non-thermal emission processes could indicate the presence of
acceleration of particles up to the relativistic regime, due to shocks
between the wind and the circumstellar environment, or due to the
wind-wind interaction like in symbiotic systems, or other processes.

In colling wind binary (CWB) models for WR+O systems, the turnover frequency is usually lower than 5.5 GHz \citep[e.g.][and ref. therein]{2010Dougherty}. Therefore, in the observed range of
frequencies, an hypothetical non-thermal component should be in the optically
thin regime, with a negative spectral index. The maximum flux
density is expected to be around 5.5 GHz or lower frequencies. Negative spectral
indices should be evident in the spectral index map obtained by
comparing the 5.5 and 9 GHz images. We prefer not to use the map at 17
GHz to derive spectral index maps, since, as reported in Section
\ref{sec:morphology}, artifacts can be present.

A spectral index map has been derived from the data at 5.5 and 9
GHz. For both bands, the LAS is much greater that the source and no
flux is lost. The 9 GHz map has been convolved with a 2-dimensional
Gaussian to match the beam at 5.5 GHz. After regridding the two maps in
order to have the same pixel size, we computed the spectral index map
and its associated error map (left and right panels of
Fig.~\ref{fig:spix_lbv}, respectively) in each common pixel
$>3 \sigma$. In the error map, the error in each pixel is dominated by
the thermal noise (flux calibration errors are negligible). We also
overlay the contours of the 9 GHz emission on top of the spectral
index map.  The mean spectral index over the nebula $\sim0.0$ is still
consistent with optically thin free-free emission from a nebular gas
ionized by the central star. We exclude in our analysis the pixels at
the borders, where the errors are high (up to 0.5). Around the central
star $\alpha\approx 0.6$--$0.7$, which is consistent with
an ionized wind. Along the diagonal arm, that is at $\rm PA=70^\circ$,
$\alpha\approx -0.2$--$0.2$ which is consistent with a typical
bremsstrahlung emission.  Near the Northern and Southern caps we find
similar values of $\alpha$, even if the associated error is much
larger there. There is no evidence of a non-thermal component, at
least at the resolution and sensitivity achieved by these
observations.

\subsection{The Spectral Energy Distribution from the near-IR to the radio}
\label{sec:dust}
\begin{figure*}
  \begin{minipage}{1\linewidth}
    \centering
    \includegraphics[width=0.8\textwidth]{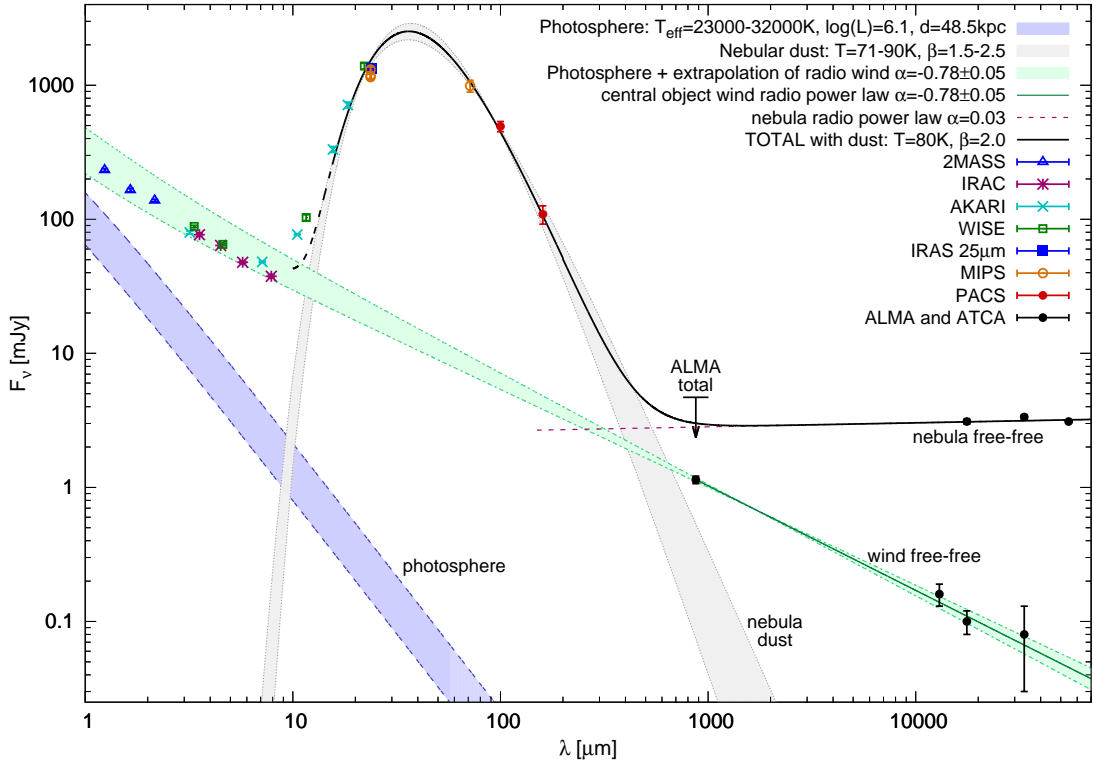}
  \end{minipage}
  \caption{SED of \rmc from near-IR to radio, including the photometry
    extracted from the IR catalogs of space telescopes and our ALMA
    and ATCA measurements. The arrow is the 3$\sigma$ upper limit to
    detect the whole nebula (dust + ionized gas) with ALMA. The gray
    band represents the fit gray-body functions obtained with a range
    of values for the parameter $\beta$ (between 1.5 and 2.5). The
    blue band represents several stellar photospheres derived assuming
    a stellar luminosity of $10^{6.1}\,\rm L\odot$ and an effective
    temperature range of $23000-32000$ K. The purple 
    line is the fit of the radio nebula and the green line the fit
    of the central object, from the centimeter to the
    sub-millimeter. The green band was derived by taking into account
    the uncertainty of the radio spectral index. The wind free-free
    fit was extrapolated up to the near-IR wavelengths and summed with
    the photosphere emission. The black continuous line is the total
    emission from the mid-IR to the radio.}
  \label{fig:sed}
\end{figure*} 

We queried the IR catalogs with the VizieR tool
\citep{2000Ochsenbein}, and we extracted the flux densities of \rmc
from 2MASS \citep{2003Cutri}, \textsl{Spitzer/IRAC}
\citep{2006Meixner}, \textsl{AKARI}
\citep{2010aIshihara,2010bIshihara}, \textsl{WISE} \citep{2012Cutri},
\textsl{IRAS} \citep{1988iras}, \textsl{Spitzer/MIPS}
\citep{2008Whitney,2011vanAarle}, and \textsl{Herschel}
\citep{2013Meixner}. In Fig.~\ref{fig:sed} we plot the spectral energy
distribution (SED) of \rmc from the near-IR to radio wavelengths. In
addition to the two power-laws associated with the ionized nebula and
with the stellar wind in the radio and sub-mm, it is also possible to
recognize a component of cool dust commensurate with a gray-body. We
also note that the photometry from about 1 to 8 $\rm{\mu m}$ traces
neither a hot dust component close to the star nor cool dust in the
outer nebula. The near-IR emission also shows an excess above the
stellar photosphere \citep[here we plot a range of reasonable
effective temperatures for \rmc during its decline toward the
quiescent state, e.g.][]{1983Stahl, 2008Walborn}. Instead, the
extrapolation of the stellar wind fit determined in the radio-mm range
seems to account for the emission in the near-IR.

We fit the SED from the mid- to the far-IR with a single-temperature gray-body with power-law opacity index $\beta$. The slope in the Rayleigh-Jeans regime suggests high values for the
parameter $\beta$, implying a grain size distribution dominated by
small grains, similar to interstellar dust. The parameter $\beta$ is
mostly constrained by the \textsl{Herschel} PACS photometry. We found
a range of characteristic temperatures between $71$ to $90\,\rm K$ by
varying $\beta$ between 1.5, 2 and 2.5 (extreme case). The gray-bodies
that fit the data, taking into account their uncertainty, are
represented in gray. The gray-body that best fits the data
is plotted with a black solid line in Fig.~\ref{fig:sed}, with
$\beta=2.0$. At longer wavelength the black solid line is summed with
the nebula free-free model, while at shorter wavelengths the total
emission is not computed because of uncertainty of the wind spectral
index and of the stellar effective temperature (green + blue
bands). Furthermore, the mid-IR range around $10\,\rm \mu m$ is known
to be complicated by solid state features that we cannot constrain.

The resulting characteristic temperature suggests that the mid- to
far-IR emission arises from optically thin cool dust in the outer
nebula \citep[consistent with][]{2009Bonanos}. In the plot, the point
indicated by the ``ALMA total'' label represents the 3$\sigma$ upper
limit to detect the total emission over the nebula (the upper limit is
derived from the rms noise in the maps integrated over the area
corresponding the ionized nebula). The point-source detected with ALMA
(black point) is clearly associated with the ionized gas in the
stellar wind (Sec. \ref{sec:fluxdensity}).

\section{The central object: discussion}

The positive slope ($\alpha$=$0.78\pm0.05$) of the radio flux density
distribution (Sec. \ref{sec:fluxdensity}) indicates a thermal origin,
so the emission must be associated with free-free encounters in the
ionized stellar wind. This value deviates from the canonical case of a
spherical wind with $n_{e}(r)\propto r^{-2}$
\citep[$\alpha$=0.6,][]{1975Panagia,1975W&B}. The spherical wind model
requires an electron density distribution with a power-law steeper
than -2 to reproduce such a spectral index.

None of the clumpy stellar wind models can reproduce the observed
radio SED. In fact, optically thin clumps (micro-clumping case) do not
alter the flux density distribution of the stellar wind at radio
wavelengths \citep{1998Nugis}. \citet{2016Ignace} recently showed that
porous stellar winds (optically thick, macro-clumping) have a spectral
index of $\nu^{0.6}$ if the porosity is in the form of shell-fragments
(for any value of volume filling factor). If the clumps are spherical,
and for extreme values of the filling factor, the flux density
distribution can be shallower than $\nu^{0.6}$ and therefore produce
an opposite effect to the \rmc case.
 
\citet{2016Daley} investigated the contribution due to the stellar
wind acceleration region in the sub-mm, but they considered stars with
relatively low mass-loss rates and with physical properties different
from LBVs. The acceleration of the wind in \rmc very likely occurs
much deeper in the wind, as indicated by the 2MASS points (see
Fig. \ref{fig:sed}).

We recall that \citet{1993Schulte} and \citet{2005Davies} found strong evidence of asphericity in the \rmc stellar wind, by means of optical spectro-polarimetry. \citet{1995Clampin} and \citet{2003Weis} also suggested a deviation from spherical symmetry by morphological considerations of the outer nebula. This is also confirmed in the radio by our new interferometric maps. These results make unsuitable all the models based on spherical symmetry. As an alternative, we employ the \citet{1986Reynolds} model of a collimated ionized stellar wind to explain the central object emission of \rmc in the radio and sub-mm. Ionized collimated stellar winds (jets) can have $-0.1<\alpha<2$ \citep{1986Reynolds}. 

\label{sec:discussion}
\subsection{Collimated stellar wind models}

\begin{table}
  \centering
  \caption{Assumed and derived parameters for model 1 (isothermal, constant velocity, fully ionized conical outflow) and for model 2 (isothermal, constant velocity, well-collimated outflow with recombinations).}
  \label{tab:models}
  \begin{tabular}{lcc}
    \tableline\tableline
    & Model 1 & Model 2 \\
    \tableline\tableline
    $q_{v}$ &\multicolumn{2}{c}{0}\\[3pt]
    $q_{T}$ &\multicolumn{2}{c}{0}\\[3pt]
    $\vartheta_{0}$ &\multicolumn{2}{c}{0.5}\\[3pt]
    $q_{x}$ &0 & -0.2\\[3pt]
    $\varepsilon$ &1.34$^{+0.15}_{-0.12}$ & 1.03$^{+0.13}_{-0.10}$\\[3pt]
    $q_{n}$ &-2.68$^{+0.24}_{-0.29}$ & -2.05$^{+0.21}_{-0.26}$\\[3pt]
    $q_{\tau}$ &-4.02$^{+0.36}_{-0.44}$ & -3.48$^{+0.31}_{-0.38}$\\[3pt]
    $q_{\rm y}$&-0.52$\pm0.05$ & -0.60$\pm0.06$\\[3pt]
    F &1.02$\pm0.10$&1.18$\pm0.12$\\[3pt]
    \tableline
  \end{tabular}
\end{table}

The Reynolds model can account for different configurations of the
jet, described as power-law dependencies of the physical parameters
along the jet (coordinate $r$ along the jet axis), such as jet width
($w\propto r^\varepsilon$), velocity ($v\propto r^{q_{v}}$), degree of
ionization ($x\propto r^{q_{x}}$), temperature ($T\propto r^{q_{T}}$),
and electron density $n_\mathrm{e}\propto r^{q_{n}}$ with
$q_{n}=-q_{v}-2\varepsilon$. Assuming for the Gaunt factor
$g_\mathrm{ff}\propto \nu^{-0.1}$, the absorption coefficient is
$k_\nu\propto r^{q_\tau} $, where
$q_{\tau}=\varepsilon+2q_{x}+2q_{n}-1.35q_{T}$.

Knowing the spectral index $\alpha$ of the wind, it is possible to
determine how the jet width varies with distance (parameter
$\varepsilon$), and therefore whether the jet is well-collimated or
conical. The relationship between $\alpha$ and $\varepsilon$ is
\begin{equation}
 \alpha = 2 + \frac{2.1}{q_{\tau}}(1+\varepsilon+q_{T})
\end{equation}

In the case (Model 1) of an isothermal ($q_{T} = 0$), fully ionized
($q_{x} = 0$) and constant velocity outflow ($q_{v} = 0$), with
$\alpha$=$0.78\pm0.05$ derived from the observations,
$\varepsilon \approx 1.34$ and then, being $>1$, the jet opens toward
the outside. Another interesting case (Model 2) to consider is that
of an isothermal, constant velocity, exactly conical
($\varepsilon = 1$) outflow with increasing recombinations ($q_{x}=-0.2$) as the
plasma propagates outwards. 

The mass-loss rate can be written in a general
form that takes into account all these parameters \citep[for details see][]{1986Reynolds}:
\begin{align}
   &\dot{M}=5.27\times 10^{-9}\, v_{\infty\, \rm [km\,s^{-1}]}\, S_{\rm[mJy]}^{3/4}\, D_{\rm [kpc]}^{3/2}\, \nu_{\rm[GHz]}^{-3/4\alpha}\, T_{\rm [K]}^{-0.075}\times \nonumber \\
   &\quad \quad \times\nu_{max\rm[GHz]}^{3/4\alpha-0.45}\, \frac{\mu}{x_{0}}\, \vartheta_{0}^{3/4}\, (\sin(\varphi))^{-1/4}\, F^{-3/4} \,[\rm M_\odot\,yr^{-1}],
	\label{eq:mass-loss}
\end{align}
where
$F\equiv F(q_{\tau}, \alpha)\equiv
\frac{(2.1)^{2}}{q_{\tau}(\alpha-2)(\alpha+0.1)}$ .

In the equation we use the flux density $S$ at frequency
$\nu = 349.5\,\rm GHz$ and 48.5 kpc as distance $D$ of the object. We
assumed $6420\pm300\,\rm K$ for the gas temperature $T$
\citep{1998Smith}, whose influence on the mass-loss rate is very
weak. For the terminal velocity of the wind we adopt
$v_{\infty}=148\pm14\,\rm km\,s^{-1}$ \citep{2017AgliozzoB}. The angle $\varphi = 90^{\circ} -i$ formed between the jet
axis and the line of sight is a free parameter, which only weakly
affects the result due to the $-1/4$ power dependence. Here we assume
$\varphi=75\,\rm \deg$ (thus $i=15\,\rm \deg$). Finally, we set to
unity the ionized fraction $x_{0}$ at the base of the outflow and the
mean atomic weight of the gas $\mu=1$ (assuming a gas of mostly
hydrogen). Note that we do not know the jet opening angle
$\vartheta_{0}$. An upper limit can be set equal to $0.5\,\rm rad$, a
condition usually met in ionized jets \citep{1985Mundt,1986Reynolds}.

Another free parameter in Reynolds' treatment is the maximum frequency
$\nu_{max}$ in the SED, that we do not know. In fact, looking at
Fig.~\ref{fig:sed}, the extrapolation of the wind SED from the radio and sub-mm wavelengths seems to approach the emission of the star in the near-IR. However, it is important
to note that the dependence of the mass-loss on $\nu_{max}$ is weak,
since it goes as $\dot{M}\propto \nu_{max}^{0.135}$ in our case. A
factor 100 in $\nu_{max}$ corresponds to a factor less than 2 in
$\dot{M}$. It is further important to note that the Reynolds equations
are based on the approximation of the Gaunt factor that is valid in
the radio regime, and that at $\nu > 10^{12}$ Hz it deviates more than
30\% from the correct value. In addition, the cutoff of the free-free
emission at high frequency, given by the factor $e^{-h\nu/KT}$, should
be taken into account when extrapolating the Reynolds equations to the
near-IR. The cutoff frequency corresponds to $10^{14}$Hz at
$T\approx6\,500$\,K, i.e. a wavelength of a few microns. Furthermore,
near the photosphere the wind is accelerated and it should be taken
into account in the model. A reasonable value for $\nu_{max}$ seems to be $10^{13}$
Hz. For the two model scenarios we then have,
\begin{align*}
   &\dot{M}_{1,2} = C_{1,2}\, \left(\frac{D}{48.5\,\rm kpc}\right)^{\!1.5}\,\frac{v_{\infty}}{148\, \rm km\,s^{-1}}\, \left(\frac{\vartheta_{0}}{0.5}\right)^{\!0.75}  \mu\, \times \\
   &\quad \quad \times \left(\frac{\nu_{max}}{10^{4}\,\rm GHz}\right)^{\!0.135}\,\sin(\varphi)^{-0.25}\, \times 10^{-6}\,[\rm M_\odot\,yr^{-1}] \\
\end{align*}
with the only difference being the normalizations
$C_1 = 9.4^{+2.6}_{-2.0}$ and $C_2 = 8.5^{+2.3}_{-1.8}$. 

The mass-loss rate can be a factor of two or more smaller than in the
spherical case ($2.1\pm0.4\times10^{-5} \,\rm M\odot\,yr^{-1}$), as
deduced from the equation of \citet{1975Panagia}.  As described in
\citet{1986Reynolds}, the effect of collimated winds is to reproduce
the radio flux density very efficiently, despite lower mass-loss rates
than in the standard spherical case. This means that for unresolved
radio stellar objects their mass-loss rates can be overestimated if
the wind is not spherical.

Astrophysical objects that exhibit jets are usually associated with
fast rotation and/or dense disks \citep[e.g.][]{1994Soker, 2000Livio}. We do not have evidence of a dense
disk in our data \citep[although][suggested the presence of a disk at
a few stellar radii]{1993Schulte,2005Davies}, but we also do know that
\rmc is a fast rotator \citep[with a projected rotational velocity of
$\sim105\,\rm km\,s^{-1}$,][]{2017AgliozzoB}. A
collimated outflow was discovered from the evolved B[e] star MWC137
\citep{2016Mehner}.

\section{The outer nebula: discussion}
\label{sec:outernebula}
The nebula associated with \rmc consists of dust and ionized gas, typical of LBVNe. In the radio the nebula emits mainly by free-free transitions (Sec. \ref{sec:spix}). 
The dust is very likely dominated by small grains, spread out over the ionized region, with an average temperature of $\rm T=80\pm10\, K$ (Sec. \ref{sec:dust}). 
Using the flux densities extracted from the fits in Sec. \ref{sec:dust} at the ALMA frequency \hbox{343.5\,GHz} (see
Fig.~\ref{fig:sed}), we derived a dust mass range of $M_{d} = 2.2\times10^{-3}$
to $2.2\times10^{-2}\,\rm M\odot$, considering that
$M_{d}=S_{\nu}\,D^{2}/(B_{\nu}(T)\,\kappa_{\nu})$, and assuming a $\kappa_{343\rm \,GHz}=1.7 \rm \,cm^{2}\,g^{-1}$, as in Paper II for \suno. The range of dust masses in \rmc's nebula is consistent with typical values in LBVs, but suggests a lack of dust when compared to the \rmc Galactic twin, AG Car \citep[e.g.][]{2015Vamvatira}, although AG Car's distance has been recently questioned \citep{2017Smith}. A reduced dust mass of \rmc compared to AG Car could be due to the lower LMC metallicity.

\paragraph{The asymmetric expanding shell}
According to the canonical view, \rmc's nebula is an expanding shell formed through past mass-loss events. The shell is not perfectly spherical and has an elongation in the N-S direction \citep{2003Weis}. The Northern and Southern Caps are also visible in the radio maps, especially at 5.5, 9 and 17 GHz (Fig. \ref{fig:maps}). 

The cause of this asymmetry could have been a dense disk in the rotational plane of the star \citep[nearly E-W direction]{1993Schulte}. This disk channeled the wind along the polar axis and expanded more slowly than the ejecta at the higher stellar latitudes, causing a density anisotropy in the nebula. The radio emission co-spatial with the optical Eastern and Western Rims would be brighter because here the optical depth along the line of sight is larger. According to this scenario, closer to the star, there would be a similar system (consisting of a dense disk and a bipolar outflow) aligned with the outer nebula.
 
This scenario is akin to B[e] supergiants, which are fast rotators and
have a dense disk in their equatorial plane and a fast outflow along
the polar axis.

\paragraph{The precessing jet model}
 
\begin{figure*}
  \centering
  \epsscale{.80}
  \plotone{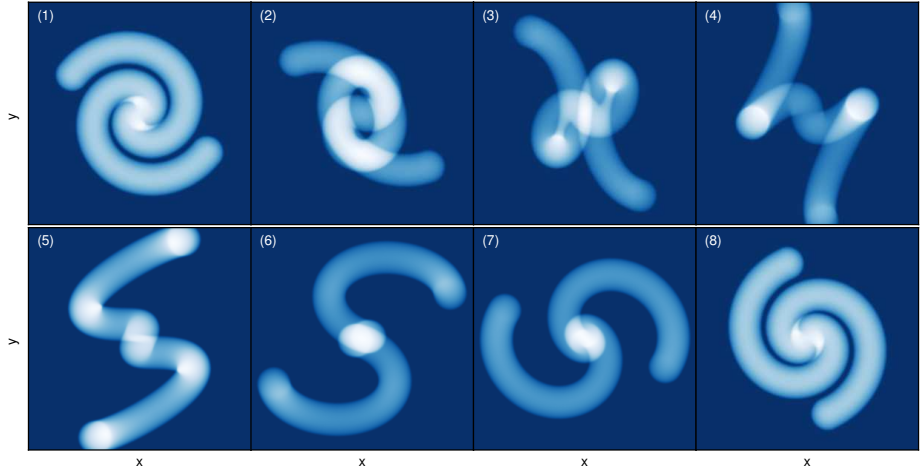}
  \caption{Simulated 3-D conical helix (projected), as seen from eight
    different lines of sight (LOS). The viewing angles between the LOS
    and the three axes $(xyz)$ change continuously in panels (1)
    through (8). Each half-helix completed exactly one turn
    ($2\pi$). The helix ``tube'' diameter and height from origin along
    the system axis are 0.15 and 0.7, respectively, in relative
    units. The helix radius at any given height is $r(h) = h$. The
    simulation was created by using the new public code RHOCUBE
    \citep{2016N&A,2017AgliozzoA}.}
  \label{fig:helix}
\end{figure*}
It is widely accepted that two thirds of massive stars are in binary
or multiple systems. The interest in the effect of binarity on the
evolution of massive stars has been increasing in recent years
\citep[e.g.][and references therein]{2017Demarco}. Recently a
companion star of the Galactic LBV HR Car was directly detected
\citep{2016Boffin}. On the basis of this discovery, \citet{2017Buemi}
suggested a precessing jet model that can explain in part the complex
HR Car nebular morphology.  HR Car's nebula is in fact characterized
by an infrared inner shell and, in the perpendicular direction, a
bipolar outflow of ionized gas, that resembles a helix. These features
would be created under the influence of the binary or multiple
system. Similarly to HR Car, \rmc has a collimated stellar wind. \rmc
is also a fast rotator. \rmc's nebula could also have been formed
through the bipolar outflow of a precessing star. To investigate this
we used our new public code RHOCUBE \citep{2016N&A,2017AgliozzoA} to
simulate a 3-D double conical helix nebula. Fig.~\ref{fig:helix}
illustrates this simulated nebula as seen along eight different lines
of sight. The three viewing angles w.r.t. the observer change
continuously between panels (1) and (8) of the figure\footnote{An
  animated version can be found at:
  \url{https://vimeo.com/151528747.}}. We find a remarkable similarity
of panel (2) in the figure with the 17-GHz morphology of \rmc. The
simulation was performed by assuming that the axial precession of the
star has completed one period and that this corresponds to the
kinematical age of the nebula. In this scenario, the polar axis is
nearly in the E-W direction and is consistent with the first
hypothesis by \citet{1993Schulte}, that the polarized emission could
arise from a highly inclined bipolar outflow. The simulation in the
figure does not have any quantitative relevance and is only shown to
provide to the reader a schematic idea of the proposed
scenario. Integral field unit spectroscopy in the future will allow to
test this geometry. In the following, we proceed with a toy model to
explore the possible implications and to eventually demonstrate that
the hypothesis of the conical helix nebula is plausible.

\paragraph{The binary toy model}
In the hypothesis that the jet precession toy model is valid, from
kinematical considerations we can derive the precession period, but
this requires an assumption on the velocity field in the outer nebula.
\citet{2003Weis} found an average projected velocity of
$25\,\rm km\,s^{-1}$ along the two rims and of about
$15\,\rm km\,s^{-1}$ along the two caps. For simplicity, we analyze
the case of a jet expanding at constant velocity \citep[set equal to the
terminal velocity of the wind, $148\pm14\,\rm km\,s^{-1}$,][]{2017AgliozzoB}. We consider then the size of the diagonal arm,
$2.8\pm0.4\,\rm arcsec$ (labeled ``a'' in Fig.~\ref{fig:maps}), and
derive a period of $4300\pm700\,\rm yr$.

The axial precession motion would imply that the star must experience
a tidal force, a torque of a companion star. Two other well-known LBVs
in a binary and a multiple system are also bright in the X-rays. These
are $\eta$ Car and HD5980 \citep{1995Corcoran, 2002Naze}. The class of
LBVs is not overall intrinsically bright in the X-rays and the known
X-ray emitters (in total four objects plus two candidates in the
Galaxy, and one in the Small Magellanic Cloud) must be generally
associated with an external factor, such as binarity
\citep{2012Naze}. Following the analysis of \citet{2012Naze}, a
massive companion O-type star, that is a moderate X-ray emitter
\citep[$L_{X}\sim10^{31}-10^{33}\,\rm erg\,s^{-1}$,][]{2012Naze},
could be invisible at the X-ray wavelengths because of the strong
absorption of the dense LBV wind, in the case of close
orbits. However, wind-wind collisions should produce
X-rays. Conversely, if the orbit is large, the intrinsic X-ray
emission associated with the O-star would be visible, while the
wind-wind interaction should not produce X-rays. A late-B companion of
3-6$\,\rm M\odot$ would not produce X-ray bright colliding wind
emission, because its wind is negligible and would be invisible at
X-ray wavelengths \citep{2010Kashi}.  We searched the X-ray archives
of $XMM-Newton$, $Chandra$ and $Swift$. One 28ks $XMM-Newton$
observation included \rmc, however no X-ray photons are detected from
its position. Assuming the distance of the LMC and the 1$\sigma$
sensitivity in the archive, the upper limit of X-ray luminosity in the
0.2-2.0 keV energy range is
$L_{X}\lesssim3\times10^{34}\,\rm erg\,s^{-1}$ and in the 2.0-12.0 keV
range is $L_{X}\lesssim1\times10^{35}\,\rm erg\,s^{-1}$. The X-ray
observations did not reach the necessary sensitivity to detect sources
as bright as the Magellanic system HD5980 (LBV+WR+O), which has X-ray
luminosities of $L_{X}=1.7\times10^{34}\,\rm erg\,s^{-1}$ in the range
0.3-10 keV and $L_{X}=9\times10^{33}\,\rm erg\,s^{-1}$ in the range
0.2-2.4 keV \citep{2002Naze}.

We analyze the case of an intermediate B-type companion for \rmc, with
mass $M_{2}=12\,\rm M\odot$, equivalent to a mass-ratio $q$ of 0.2,
typical in observations of massive binary systems in our Galaxy
\citep{2014Kobulnicky}.  For the companion to exert a torque, \rmc
must be not perfectly spherical and its equatorial plane must lie at
an angle with the plane of the orbit.  The angular velocity of the
precession axis is $\Omega_{p} = \tau / I \Omega_{r} \sin(\theta)$,
where $\theta$ is the angle formed by the rotational axis with the
precession axis, $I$ the moment of inertia, $\Omega_{r}$ the angular
velocity of the stellar axis. The magnitude of the torque is then
$\tau=R_{*} \Delta F \sin(\theta)$, where $R_{*}$ is the stellar
radius, $\Delta F$ is the gravitational force across the star's width
and is $\Delta F = 2 G M_{1} M_{2}\Delta r / a^3$, with $M_{1}$ and
$M_{2}$ the masses of the two stars in the binary system and $G$ the
gravitational constant.  Therefore, the linear separation between the
two stars $a$ is

\begin{equation}
  a=\left(4\frac{G M_1 M_2 R_{*}^2 \cos(\theta)}{I \Omega_{r} \Omega_{p}}\right)^{1/3}
\end{equation}

To estimate the angular velocity $\Omega_{r}$ of \rmc we take the
projected rotational velocity of $105\,\rm km\,s^{-1}$ \citep{2017AgliozzoB} and the projection angle $i = 15\,\rm \deg$
(consistent with our model in Section \ref{sec:discussion}). If we
assume as stellar radius $R_{*}$ = 50 R$_{\sun}$ and mass $M_{1}$ = 60
M$_{\sun}$ \citep{1983Stahl}, we obtain
$\Omega_{r}\sim1.2\times10^{-5}\, \rm rad\,s^{-1}$, which corresponds
to a period of $\sim6$ terrestrial days. For the moment of inertia we
approximate a solid sphere, given the small dependence of $a$ on $I$. From the precession period
$T_{p}\simeq4300\,\rm yr$ we derive
$\Omega_{p}\sim4.6\times10^{-11}\,\rm rad \,s^{-1}$. Finally, based on
the similarity between the simulation in Fig.~\ref{fig:helix} and the
map at 17 GHz, we assume $\theta=45^{\circ}$.

In this examined case (companion of 12 M$_{\sun}$) the inner separation $a$ between the two stars would be then $\sim18\,\rm AU$ and the inner Lagrangian point L1 would be $\sim12\,\rm AU$ ($\sim51\,\rm R_{*}$), implying that the system is detached. However, when the primary star is at its maximum phase, the expanded pseudo-photosphere \citep[stellar radius of $150\,\rm R_{\odot}$,][]{1983Stahl} would fill the Roche lobe, implying mass transfer. The orbital period for this particular orbit would be about 9 yr. %
\citet{2016Boffin} derived for HR Car's binary system a linear separation of 18 AU, an orbital period of 12 yr and a mass-ratio of 0.36. 
The presented exercise shows that the hypothesis of binarity for \rmc and axial precession is reasonable.

Noticeably, \cite{2016Lau} found an apparent precessing helical outflow associated with the Wolf Rayet star WR102c and attributed it to a previous phase of its evolution (namely, LBV). They also concluded that the helix is evidence of a binary interaction. They derived a precession period of 14000 yr. 

The precessing jet model depends on the assumption of the binary nature of \rmc, which has not yet been demonstrated.  
Given this, the single star expanding shell scenario appears the simplest description for the nebula. 
A long term, multi-wavelength observation campaign will be needed to conclusively distinguish these two scenarios and understand the nature of this complex object.

\section{Summary}
\label{sec:summary}

The ALMA and ATCA observations of \rmc's central object and outer
nebula provide new insights of the nebula core of the classical LBV
\rmc. In the radio, at the lowest frequencies, the main component of
emission is the ionized gas in the outer nebula, that resembles
overall the $\rm H\alpha$ emission. The radio data permitted us to
also analyze the inner part of the nebula, which in the optical is
obscured by the bright central star. In addition to the previously
known features in the nebula (Northern and Southern caps, Eastern and Western
rims), we detected another emission component that gives to the nebula
a strongly asymmetric aspect, a Z-pattern shape. We noticed a similar
morphology in the \emph{HST} $\rm H\alpha$
image.

The emission mechanism for the outer nebula in the radio is overall
optically thin free-free with a global spectral index $\alpha$ of
$-0.03\pm0.08$. At higher frequencies a point-source component appears
at the position of the star, bright up to the ALMA observing frequency
of 349.5 GHz. This emission is due to thermal free-free emission in the
ionized stellar wind. The stellar wind also seems to account for the
excess at the near-IR wavelengths above the photosphere. The flux
density distribution of the ionized wind (with spectral index
$\alpha$=0.78$\pm$0.05) indicates a deviation from a spherical wind,
supporting previous studies, and likely suggests the presence of a
bi-polar outflow/jet. We fitted the data with two \citet{1986Reynolds}
models to determine the mass-loss rate in the jet, which can be at
least a factor of two smaller than the case of spherical wind.

 The fit of the mid- to far-IR flux densities derived from space telescope observations suggests that this emission arises from optically thin cool ($\sim80\,\rm K$) dust spread out over the ionized region. The derived mass of the dust ($0.2-2.2\times10^{-2}\,\rm M\odot$) is consistent with other Magellanic and Galactic LBVs. 

 We discussed two possible geometries to explain the outer nebula, including the canonical single star expanding shell model and a jet precession model assuming the presence of a companion star.
 
The asymmetry of the mass-loss geometry
of \rmc may be strongly influenced by fast rotation and/or the presence of a companion star.

\acknowledgments

We are thankful to Yazam Momany for assistance with the VISIR data. We thank the referee for thorough feedback which has helped to improve and
clarify our presentation of this work. 
We acknowledge support from FONDECYT grant No. 3150463 (CA), FONDECYT grant
  No. 3140436 (RN), FONDECYT Regular 1141218 (FEB), FONDECYT grant 1151445 (JLP), 
CONICYT-Chile grants Basal-CATA PFB-06/2007 (FEB), 
"EMBIGGEN" Anillo ACT1101 (FEB), and
the Ministry of Economy, Development, and Tourism's Millennium Science
Initiative through grant IC120009, awarded to The Millennium Institute
of Astrophysics, MAS (CA, GP, JLP, FEB). We also wish to thank the staff at ESO, ALMA and ATCA who made these observations possible.

This paper makes use of the following ALMA data:
\dataset{ADS/JAO.ALMA\#2013.1.00450.S}. ALMA is a partnership of ESO
(representing its member states), NSF (USA) and NINS (Japan), together
with NRC (Canada) and NSC and ASIAA (Taiwan) and KASI (Republic of
Korea), in cooperation with the Republic of Chile. The Joint ALMA
Observatory is operated by ESO, AUI/NRAO and NAOJ. This paper also
includes: data collected at the European Organisation for Astronomical
Research in the Southern Hemisphere under ESO programmes 095.D-0433(A)
and 095.D-0433(B).  The Australia Telescope Compact Array is part of
the Australia Telescope National Facility which is funded by the
Australian Government for operation as a National Facility managed by
CSIRO. This work made use of PyAstronomy.

{\it Facilities:} \facility{ATCA}, \facility{ALMA}, \facility{VLT}.



\appendix

\section{Additional notes on the ALMA observation and data reduction}

A standard Band 7 continuum spectral setup was used with the 64-input Baseline Correlator, giving four 2 GHz-width spectral windows (one per analogue baseband) of 128 channels ("TDM" mode, XX+YY polarization correlations) centered at approximately 336.5(LSB), 338.5(LSB), 348.5(USB) and 350.5(USB) GHz, with integration duration of 2.016 seconds. Companion channel-averaged correlator data with integration duration 1.008 second, and Water Vapor Radiometer (WVR) data with integration duration 1.152 second were also recorded. Time on source was approximately 16 minutes per target. Atmospheric conditions were marginal for the combination of frequency and necessarily high airmass (transit elevation $43^\circ$ for \rmc), requiring extra calibration steps described below.

Of the 40 antennas, two had to be completely flagged (DA53, DV06), and another flagged completely in three of the four basebands (DA49 BB\_2,3,4) due to intermittent coherence loss (a digitizer calibration problem affecting Walsh sequence phase switching). For one antenna (DV11) manual intervention was required in order to produce system temperature measurements (intermittent spurious values in the calibration device load temperature data). System temperatures were re-generated offline using the Cycle-3 {\tt TelCal} software. Flags set by the online control software (XML flags) and by the correlator software (binary data flags) were applied as normal. In total 36 antennas were fully used in the reduction, with two more partially used due to issues in a subset of BBs/pols (DA49 BB\_2,3,4; DA45 pol Y).

Online, antenna focus was calibrated in an immediately preceding execution, and antenna pointing was calibrated on each calibrator source during the execution (all using Band 7). Scans at the science target tuning on bright quasar calibrators \bpcal and \fluxcal (PKS J0519-4546; an ALMA secondary flux calibrator `grid' source) were used for interferometric bandpass and absolute flux scale calibration. Astronomical calibration of complex gain variation was made using scans on quasar calibrator \qso interleaved with scans on the science targets approximately every six minutes. The gain calibrator was a sub-optimal choice, as being six degrees further South than the targets it was at significantly higher airmass, with many antennas suffering some degree of shadowing. Data reduction proceeded as normal for ALMA data reduced in CASA, with the addition of the following modifications to deal with the combination of large airmass separation between science targets and gain calibrator, shadowing of antennas due to the compact configuration and low observing elevation, and generally marginal phase stability. We also evaluated the effect of calibrator source structure on the calibration.

\subsubsection{Continuum WVR subtraction}

Before running the {\tt wvrgcal} program \citep{2012Nikolic} which computes phase corrections from the WVR data, we pre-processed the raw WVR data to subtract a continuum contribution using a prototype algorithm implementation developed at JAO (W. Dent, priv. com.). This was developed to subtract the thermal continuum contribution produced by water droplets from the WVR channel temperatures, as {\tt wvrgcal} assumes only water \emph{vapor} emission. In this case it was primarily used to remove the thermal continuum due to shadowing (partially obstructed beam) from the WVR data, with the same reasoning. Reviewing the corrections applied to each antenna, compared with their predicted shadowing fraction for each scan, showed that this was successful, although in future the correction may improve by use of measured sky coupling efficiencies of each antenna+WVR combination (a topic of active investigation within the ALMA project).

\subsubsection{Removal of WVR phase offsets between fields}

Due to the large airmass separation between the science targets and the gain calibrator, combined with limitations in the calibration of the WVR data (a fixed sky coupling efficiency and channel frequencies are currently assumed) and limitations in the atmosphere model used to derive the phase corrections from the WVR data, we found phase offsets between fields in the phase correction table produced by {\tt wvrgcal}, which differed between antennas and did not correspond to real phase offsets (confirmed by looking at self-calibration phase solutions for phase over all time on \rmc -- discrepant antennas corresponded to those with noted field offsets in the {\tt wvrgcal} results). This effect is under investigation as part of ALMA's continuing improvements to phase correction and antenna position determination. Without action, the image smearing due to these offsets made the WVR phase correction no significant improvement over not applying the correction. A simple solution of subtracting the field-averaged phase correction from the calibration table produced by {\tt wvrgcal} was applied. This dramatically improved the image quality, resulting in over a 10\% increase of the peak flux of \rmc.

\subsubsection{$T_{\rm sys}$ extrapolation between fields}

The ALMA observation frequently measured the system temperature, $T_{\rm sys}$, at the location of the gain calibrator \qso. The standard ALMA data reduction applies this $T_{\rm sys}$ directly to the science fields, on the assumption that the difference is negligible due to proximity of the calibrator. This is a known limitation in ALMA's amplitude calibration strategy -- {\sc casa} provides simple interpolation of $T_{\rm sys}$ in time (between scans) and frequency (within each $T_{\rm sys}$ spectral window), but not yet in airmass. For the dataset considered here, the $T_{\rm sys}$ error for the science targets by simply using that of the lower elevation gain calibrator was around 8--10\%, with the error being largest for antennas which were more shadowed (larger blocking fraction) towards the gain calibrator. A simple $T_{\rm sys}$ extrapolation scheme was developed to correct this, using a simple model and the autocorrelation amplitude during each of the scans. This works for this dataset, as we used the TDM correlator mode, which produces linear autocorrelations (a quantization correction is applied in the correlator software, which cannot be applied in the higher resolution FDM mode). The channel-average autocorrelation data was used for this. The $T_{\rm sys}$ at the start and end of each scan was interpolated from the $T_{\rm sys}$ measurements on the gain calibrator using the following equation, taking an input $T_{\rm sys,1}$ and the autocorrelation values $V_{\rm 1}$, $V_{\rm 2}$ at the relevant times.
\begin{equation}
T_{\rm sys,2} = T_{\rm atm} \left(\frac{x}{1-x}\right);\quad x=\frac{T_{\rm sys,1}}{T_{\rm sys,1} + T_{\rm atm}}\frac{V_{\rm 2}}{V_{\rm 1}}
\end{equation}
A nominal atmosphere and blocking temperature $T_{\rm atm}=270\,{\rm K}$ was used, although the effect of varying this by plausible amounts was negligible for this case of $T_{\rm sys}\!\sim\!200\,{\rm K}$.

\subsubsection{Source structure in gain calibrator \qso}

\begin{figure}
\centerline{\includegraphics[scale=0.7]{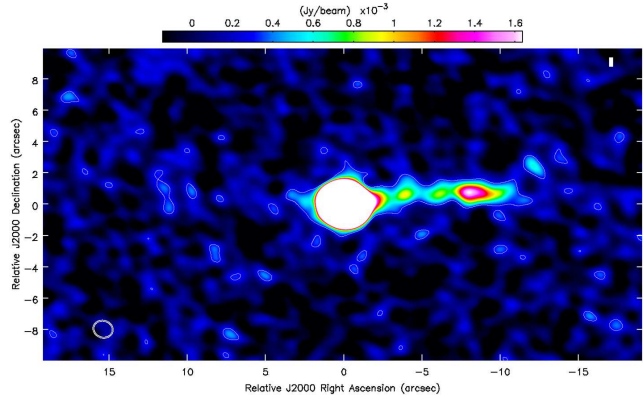}}
\figcaption{ALMA Band 7 image ($343.5\,{\rm GHz}$, natural weighting, not primary beam corrected) of gain calibrator \qso, showing source structure in the jet to the West. The image is noise-limited with RMS noise of $0.15\,{\rm mJy} {\,\rm beam^{-1}}$, which is $1/3200$ of the central source. A single contour at $+2\sigma$ is overlaid. Synthesized beam size is $1.1''$.}
\label{fig:phasecal}
\end{figure}

We imaged the three calibrator sources in the execution as a cross-check of calibration and data quality. \bpcal and \fluxcal were point sources at the expected position. The gain calibrator, \qso, however showed significant source structure as shown in Fig.~\ref{fig:phasecal}. This is a known mega-parsec scale jet discovered by {\it Chandra} \citep{J0635Chandra} and previously imaged at centimeter and optical/near-IR wavelengths \citep[e.g.][]{J0635ATCA,J0635HST}. Since analysing our ALMA observation, maps from combination of calibrator scans in many ALMA observations have been presented by \citet{2017Meyer}. To evaluate the effect of this structure on the phase calibration of the science targets, we used a {\tt clean} component model of the source to both self-calibrate it and for correcting the phases of the other fields. The maximum in the residual self-calibration phases of \qso was around $3^\circ$, and there was no significant effect on the image of \rmc, so we concluded that the source structure of \qso was irrelevant and it was a suitable calibrator choice in this regard (and it would be even less significant with smaller largest recoverable scale).

\begin{thebibliography}{}
\bibitem[Agliozzo et al.(2012)]{2012Agliozzo} Agliozzo, C., Umana, 
G., Trigilio, C., et al.\ 2012, \mnras, (Paper I), 426, 181 
\bibitem[Agliozzo et al.(2014)]{2014Agliozzo} Agliozzo, C., Noriega-Crespo, A., Umana, G., et al.\ 2014, \mnras, 440, 1391 
\bibitem[Agliozzo et al.(2017a)]{2017AgliozzoA} Agliozzo, C., Nikutta, R., Pignata, G., et al.\ 2017a, \mnras, (Paper II), 466, 213 
\bibitem[Agliozzo et al.(2017b)]{2017AgliozzoB} Agliozzo, C., et al.\ 2017b, conference proceeding ``The mass-loss before the end: two luminous blue variables with a collimated stellar wind'', IAU-17-IAUS329, in press
\bibitem[Beichman et al.(1988)]{1988iras} Beichman, C.~A., Neugebauer, G., Habing, H.~J., Clegg, P.~E., \& Chester, T.~J.\ 1988, Infrared astronomical satellite (IRAS) catalogs and atlases.~Volume 1: Explanatory supplement, 1,  
\bibitem[Boffin et al.(2016)]{2016Boffin} Boffin, H.~M.~J., Rivinius, T., M{\'e}rand, A., et al.\ 2016, \aap, 593, A90 
\bibitem[Bohannan \& Walborn(1989)]{1989BW} Bohannan, B., \& Walborn, N.~R.\ 1989, \pasp, 101, 520 
\bibitem[Bonanos et al.(2009)]{2009Bonanos} Bonanos, A.~Z., Massa, D.~L., Sewilo, M., et al.\ 2009, \aj, 138, 1003 
\bibitem[Buemi et al.(2010)]{2010Buemi} Buemi, C.~S., Umana, G., Trigilio, C., Leto, P., \& Hora, J.~L.\ 2010, \apj, 721, 1404 
\bibitem[Buemi et al.(2017)]{2017Buemi} Buemi, C.~S., Trigilio, C., Leto, P., et al.\ 2017, \mnras, 465, 4147 
\bibitem[Clampin et al.(1993)]{1993Clampin} Clampin, M., Nota, A., 
Golimowski, D.~A., Leitherer, C., \& Durrance, S.~T.\ 1993, \apjl, 410, L35 
\bibitem[Clampin et al.(1995)]{1995Clampin} Clampin, M., 
Schulte-Ladbeck, R.~E., Nota, A., et al.\ 1995, \aj, 110, 251 
\bibitem[Corcoran et al.(1995)]{1995Corcoran} Corcoran, M.~F., Rawley, G.~L., Swank, J.~H., \& Petre, R.\ 1995, \apjl, 445, L121 
\bibitem[Cutri et al.(2003)]{2003Cutri} Cutri, R.~M., Skrutskie, M.~F., van Dyk, S., et al.\ 2003, ``The IRSA 2MASS All-Sky Point Source Catalog, NASA/IPAC Infrared Science Archive. ''
\bibitem[Cutri et al.(2012)]{2012Cutri} Cutri, R.~M., Skrutskie, M.~F., van Dyk, S., et al.\ 2012, VizieR Online Data Catalog, 2281,  
\bibitem[Daley-Yates et al.(2016)]{2016Daley} Daley-Yates, S., Stevens, I.~R., \& Crossland, T.~D.\ 2016, \mnras, 463, 2735 
\bibitem[Davies et 
al.(2005)]{2005Davies} Davies, B., Oudmaijer, R.~D., \& Vink, J.~S.\ 2005, \aap, 439, 1107 
\bibitem[De Marco \& Izzard(2017)]{2017Demarco} De Marco, O., \& Izzard, R.~G.\ 2017, PASA, 34, e001 
\bibitem[Dougherty(2010)]{2010Dougherty} Dougherty, S.~M.\ 2010, High Energy Phenomena in Massive Stars, 422, 166 
\bibitem[Godfrey et al. (2012)]{J0635ATCA} Godfrey L.~E.~H. et al. 2012, \apj, 758L, 27
\bibitem[Groh et al.(2006)]{2006Groh} Groh, J.~H., Hillier, 
D.~J., \& Damineli, A.\ 2006, \apjl, 638, L33 
\bibitem[Gvaramadze et al.(2015)]{2015Gv} Gvaramadze, V.~V., Kniazev, A.~Y., Bestenlehner, J.~M., et al.\ 2015, \mnras, 454, 219 
\bibitem[Higgs et al.(1994)]{1994Higgs} Higgs, L.~A., Wendker, H.~J., \& Landecker, T.~L.\ 1994, \aap, 291, 295 
\bibitem[Humphreys \& Davidson(1994)]{1994HD} Humphreys, R.~M., \& Davidson, K.\ 1994, \pasp, 106, 1025 
\bibitem[Humphreys et al.(2014)]{2014Humphreys} Humphreys, R.~M., 
Weis, K., Davidson, K., Bomans, D.~J., \& Burggraf, B.\ 2014, \apj, 790, 48 
\bibitem[Humphreys et al.(2016)]{2016Humphreys} Humphreys, R.~M., Weis, K., Davidson, K., \& Gordon, M.~S.\ 2016, \apj, 825, 64 
\bibitem[Ignace(2016)]{2016Ignace} Ignace, R.\ 2016, \mnras, 457, 4123 
\bibitem[Ishihara et al.(2010a)]{2010aIshihara} Ishihara, D., Onaka, T., Kataza, H., et al.\ 2010, \aap, 514, A1 
\bibitem[Ishihara et al.(2010b)]{2010bIshihara} Ishihara, D., Onaka, T., Kataza, H., et al.\ 2010, VizieR Online Data Catalog, 2297,  
\bibitem[Kashi(2010)]{2010Kashi} Kashi, A.\ 2010, \mnras, 405, 1924 
\bibitem[Kobulnicky et al.(2014)]{2014Kobulnicky} Kobulnicky, H.~A., Kiminki, D.~C., Lundquist, M.~J., et al.\ 2014, \apjs, 213, 34 
\bibitem[Lau et al.(2016)]{2016Lau} Lau, R.~M., Hankins, M.~J., Herter, T.~L., et al.\ 2016, \apj, 818, 117 
\bibitem[Leitherer et al.(1994)]{1994Leitherer} Leitherer, C., Allen, 
R., Altner, B., et al.\ 1994, \apj, 428, 292 
\bibitem[Livio(2000)]{2000Livio} Livio, M.\ 2000, Asymmetrical Planetary Nebulae II: From Origins to Microstructures, 199, 243 
\bibitem[McMullin et al.(2007)]{2007McMullin} McMullin, J.~P., Waters, B., Schiebel, D., Young, W., \& Golap, K.\ 2007, Astronomical Data Analysis Software and Systems XVI, 376, 127 
\bibitem[Meixner et al.(2006)]{2006Meixner} Meixner, M., Gordon, K.~D., Indebetouw, R., et al.\ 2006, \aj, 132, 2268 
\bibitem[Meixner et al.(2013)]{2013Meixner} Meixner, M., Panuzzo, P., Roman-Duval, J., et al.\ 2013, \aj, 146, 62 
\bibitem[Mehner et al.(2016)]{2016Mehner} Mehner, A., de Wit, W.~J., Groh, J.~H., et al.\ 2016, \aap, 585, A81 
\bibitem[Mehta et al. (2009)]{J0635HST} Mehta, K.~T., Georganopoulos, M.; Perlman, E.~S.; Padgett, C.~A., Chartas, G. 2009, \apj, 690, 1706
\bibitem[Meyer et al.(2017)]{2017Meyer} Meyer, E.~T., Breiding, P., Georganopoulos, M., et al.\ 2017, \apjl, 835, L35 
\bibitem[Mundt(1985)]{1985Mundt} Mundt, R.\ 1985, Protostars and Planets II, 414 
\bibitem[Naz{\'e} et al.(2002)]{2002Naze} Naz{\'e}, Y., Hartwell, J.~M., Stevens, I.~R., et al.\ 2002, \apj, 580, 225 
\bibitem[Naz{\'e} et al.(2012)]{2012Naze} Naz{\'e}, Y., Rauw, G., \& Hutsem{\'e}kers, D.\ 2012, \aap, 538, A47 
\bibitem[Nikolic et al.(2012)]{2012Nikolic} Nikolic, B., Graves, S. F., Bolton, R. C., \& Richer, J. S. 2012, Design and Implementation of the wvrgcal Program, ALMA Memo Series 593, The ALMA Project
\bibitem[Nikutta \& Agliozzo(2016)]{2016N&A} Nikutta, R., \& Agliozzo, C.\ 2016, Astrophysics Source Code Library, ascl:1611.009 
\bibitem[Nota et al.(1995)]{1995Nota} Nota, A., Livio, M., Clampin, M., \& Schulte-Ladbeck, R.\ 1995, \apj, 448, 788 
\bibitem[Nugis et al.(1998)]{1998Nugis} Nugis, T., Crowther, P.~A., \& Willis, A.~J.\ 1998, \aap, 333, 956 
\bibitem[Ochsenbein et al.(2000)]{2000Ochsenbein} Ochsenbein, F., Bauer, P., \& Marcout, J.\ 2000, \aaps, 143, 23 
\bibitem[Panagia 
\& Felli(1975)]{1975Panagia} Panagia, N., \& Felli, M.\ 1975, \aap, 39, 1 
\bibitem[Reynolds(1986)]{1986Reynolds} Reynolds, S.~P.\ 1986, \apj, 
304, 713 
\bibitem[Sault et al.(1995)]{1995SaultMiriad} Sault, R.~J., Teuben, P.~J., \& Wright, M.~C.~H.\ 1995, Astronomical Data Analysis Software and Systems IV, 77, 433 
\bibitem[Schulte-Ladbeck et al.(1993)]{1993Schulte} 
Schulte-Ladbeck, R.~E., Leitherer, C., Clayton, G.~C., et al.\ 1993, \apj, 
407, 723
\bibitem[Schwartz et al. (2000)]{J0635Chandra} Schwartz, D.~A. et al. 2000, \apj, 540L, 69
\bibitem[Smith et al.(1998)]{1998Smith} Smith, L.~J., Nota, A., 
Pasquali, A., et al.\ 1998, \apj, 503, 278 
\bibitem[Smith \& Owocki(2006)]{2006S&O} Smith, N., \& Owocki, S.~P.\ 2006, \apjl, 645, L45 
\bibitem[Smith \& Tombleson(2015)]{2015S&T} Smith, N., \& Tombleson, R.\ 2015, \mnras, 447, 598 
\bibitem[Smith \& Stassun(2017)]{2017Smith} Smith, N., \& Stassun, K.~G.\ 2017, \aj, 153, 125 
\bibitem[Soker \& Livio(1994)]{1994Soker} Soker, N., \& Livio, M.\ 1994, \apj, 421, 219 
\bibitem[Stahl et 
al.(1983)]{1983Stahl} Stahl, O., Wolf, B., Klare, G., et al.\ 1983, \aap, 127, 49 
\bibitem[Stahl(1986)]{1986StahlO} Stahl, O.\ 1986, \aap, 164, 321
\bibitem[Umana et al.(2005)]{2005Umana} Umana, G., Buemi, C.~S., Trigilio, C., \& Leto, P.\ 2005, \aap, 437, L1 
\bibitem[Umana et al.(2011b)]{2011UmanaB} Umana, G., Buemi, C.~S., Trigilio, C., et al.\ 2011, \apjl, 739, L11 
\bibitem[Umana et al.(2012)]{2012Umana} Umana, G., Ingallinera, A., Trigilio, C., et al.\ 2012, \mnras, 427, 2975 
\bibitem[Vamvatira-Nakou et al.(2015)]{2015Vamvatira} Vamvatira-Nakou, C., Hutsem{\'e}kers, D., Royer, P., et al.\ 2015, \aap, 578, A108 
\bibitem[van Aarle et al.(2011)]{2011vanAarle} van Aarle, E., van Winckel, H., Lloyd Evans, T., et al.\ 2011, \aap, 530, A90 
\bibitem[Walborn(1972)]{1972Walborn} Walborn, N.~R.\ 1972, \aj, 77, 312 
\bibitem[Walborn(1977)]{1977Walborn} Walborn, N.~R.\ 1977, \apj, 215, 53 
\bibitem[Walborn(1982)]{1982Walborn} Walborn, N.~R.\ 1982, \apj, 
256, 452 
\bibitem[Walborn et al.(2008)]{2008Walborn} Walborn, N.~R., Stahl, 
O., Gamen, R.~C., et al.\ 2008, \apjl, 683, L33 
\bibitem[Walborn et al.(2015)]{2015Walborn} Walborn, N.~R., Morrell, N.~I., Naz{\'e}, Y., et al.\ 2015, \aj, 150, 99 
\bibitem[Wang et al.(2004)]{2004Wang} Wang, W., Liu, X.-W., Zhang, Y., \& Barlow, M.~J.\ 2004, \aap, 427, 873 
\bibitem[Weis(2003)]{2003Weis} Weis, K.\ 2003, \aap, 408, 205 
\bibitem[Weis(2011)]{2011Weis} Weis, K.\ 2011, Bulletin de la Societe Royale des Sciences de Liege, 80, 440 
\bibitem[Whitney et al.(2008)]{2008Whitney} Whitney, B.~A., Sewilo, M., Indebetouw, R., et al.\ 2008, \aj, 136, 18 
\bibitem[Wright \& Barlow(1975)]{1975W&B} Wright, A.~E., \& Barlow, M.~J.\ 1975, \mnras, 170, 41 
\bibitem[Wolf et 
al.(1988)]{1988Wolf} Wolf, B., Stahl, O., Smolinski, J., \& Casatella, A.\ 1988, \aaps, 74, 239 
\bibitem[Zickgraf et al.(1985)]{1985Zickgraf} Zickgraf, F.-J., Wolf, B., Stahl, O., Leitherer, C., \& Klare, G.\ 1985, \aap, 143, 421 



\end{thebibliography}
\end{document}